\documentclass[aps,pra,superscriptaddress,twocolumn,citeautoscript,10pt, rsi,reprint]{revtex4-2}
\usepackage{amsmath}
\usepackage{graphicx}
\usepackage{bm}
\usepackage{soul}
\usepackage{xcolor}
\usepackage{amsmath}
\usepackage{float}
\usepackage{comment}
\usepackage{graphicx}
\usepackage{dcolumn}
\usepackage[colorlinks=true,bookmarks=false,citecolor=blue,linkcolor=blue,hyperfootnotes=true,urlcolor=blue]{hyperref}

\def\GO{Ga$_2$O$_3$}
\def\BGO{$\beta$-Ga$_2$O$_3$}
\def\ALGO{(Al$_x$Ga$_{1-x}$)$_2$O$_3$}
\def\AAGO{AlGaO$_3$}
\def\AO{Al$_2$O$_3$}
\def\TAO{$\theta$-Al$_2$O$_3$}
\def\AAO{$\alpha$-Al$_2$O$_3$}
\def\AGO{$\alpha$-Ga$_2$O$_3$}

\begin{document}
\preprint{AIP/123-QED}
\title{Halide donors in monoclinic- and corundum-phase \GO\  and {\AO}}
\author{Sai Mu}\email{mus@mailbox.sc.edu}
\affiliation{Department of Physics and Astronomy and Center for Experimental Nanoscale Physics, University of South Carolina, Columbia, SC, 29208 USA}
\author{Haochen Wang}  
\affiliation{Materials Department, University of California, Santa Barbara, California, 93106, USA}
\author{Yongjoong Shin} 
\affiliation{Materials Department, University of California, Santa Barbara, California, 93106, USA}
\author{Zhi-Hao Wang}
\affiliation{Department of Physics and Astronomy and Center for Experimental Nanoscale Physics, University of South Carolina, Columbia, SC, 29208 USA} 
\author{Chris G. Van de Walle}\email{vandewalle@mrl.ucsb.edu}
\affiliation{Materials Department, University of California, Santa Barbara, California, 93106, USA}

\begin{abstract}
We present a systematic first-principles investigation of halide impurities (F and Cl) in Ga$_2$O$_3$ and Al$_2$O$_3$, considering both monoclinic and corundum phases. 
Our study of the structural properties, formation energies, and charge-state transition levels establishes the relative stability of different atomic configurations and different charge states.    
We find that F and Cl on oxygen sites act as shallow donors in Ga$_2$O$_3$, in both the monoclinic and corundum phases. However, their behavior differs substantially as the band gap increases with greater Al compositions. Fluorine is prone to $DX$-center formation with increased Al composition, leading to self-compensation at 38\% Al concentrations in monoclinic {\ALGO} and 70\% Al concentration in corundum {\ALGO}. 
Chlorine is more resistant to $DX$-center formation: in monoclinic {\ALGO}, Cl$_\mathrm{O}$ on the lowest-energy oxygen site shows an onset of $DX$ behavior at 50\% alloy compositions,
while in corundum {\ALGO}  this onset for Cl$_\mathrm{O}$ occurs only at Al concentrations as high as 84\%. 
We also study F and Cl interstitials, finding that they 
act as compensating centers but also exhibit migration barriers that are low enough for them to be removed by post-growth annealing.
Surprisingly, in both monoclinic and corundum Al$_2$O$_3$, Cl$_\mathrm{O}$ exhibits a relatively shallow transition level located at 0.48~eV below the conduction-band minimum, much shallower than F$_\mathrm{O}$ and other donor candidates. 
These remarkable results identify Cl as an unusually promising donor candidate in {\ALGO} alloys and even pure {\AO}, although high formation energies and compensation  
will render observation of true $n$-type conductivity in {\AO} difficult.

\end{abstract}
\maketitle

\section{\label{sec:level1}Introduction}

Monoclinic {\GO} ({\BGO}) is attracting a great deal of attention for power electronics~\cite{suzuki2009enhancement,oshima2008vertical,alema2019solar} due to its wide band gap (4.76---5.1 eV \cite{tippins1965optical,matsumoto1974absorption,sturm2016dipole,mock2017band}), high breakdown field (6-8 MV/cm) \cite{higashiwaki2012gallium} and the availability of high-quality yet low-cost substrates. 
Despite its wide band gap, {\BGO} can be controllably $n$-type doped, and a number of shallow donors (Si, Ge, Sn, C, H, Cl, F, Hf, Zr, Nb, W)~\cite{varley2010oxygen,lyons2014carbon, peelaers2016doping, karbasizadeh2024transition}
with modest measured ionization energies ($\le$ 80 meV) have been identified~\cite{feng2019mocvd, zhang2019mocvd,son2016electronic,parisini2016analysis,ma2016intrinsic,higashiwaki2017state,moser2017ge,neal2018donors,orita2000deep,oishi2016conduction,higashiwaki2013depletion,varley2010oxygen,varley2020prospects}. 
Additionally, as a polymorph of {\GO}, corundum {\GO}, with its higher band gap of 5.3~eV~\cite{shinohara2008heteroepitaxy,Segura2017,jinno2021crystal}, has received growing attention.
Corundum-phase {\AO} is the most stable (and technologically most relevant) polymorph of {\AO}, denoted as $\alpha$-{\AO} (sapphire).

Similar to {\BGO}, {\AGO} can also be $n$-type doped using shallow donors such as Sn and Si~\cite{dang2020conductive, polyakov2019deep, akaiwa2020electrical}.

Alloying with Al raises the band gap of {\GO}~\cite{peelaers2018structural,*peelaers2019erratum}, and
{\ALGO}/{\GO} heterojunctions that give rise to a high-density two-dimensional electron gas 
are at the heart of many devices~\cite{zhang2018demonstration}.  
Modulation doping is required for high mobility, thus raising the need for $n$-type doping of {\ALGO}.
Experiments have indicated that control of doping is challenging.
Silicon has been the most widely explored dopant.
In {\ALGO} films grown by metal-organic chemical vapor deposition (MOCVD), incorporating Si failed to result in $n$-type doping below a threshold Si concentration~\cite{bhuiyan2022si,uddin2023metalorganic}.
In bulk {\ALGO} single crystals grown by the Czochralski method, Si incorporation did not result in $n$-type conductivity above 25\% Al~\cite{galazka2023}.

Recent computational studies~\cite{varley2020prospects,mu2022role,wickramaratne2022} aimed to identify the factors that can limit doping in {\ALGO} alloys.  
These include compensation by native defects \cite{varley2011hydrogenated,ingebrigtsen2019impact} or by impurities that act as acceptors~\cite{mu2022role}, and also self-compensation~\cite{mu2022role,wickramaratne2022}.
The latter can occur if substitutional impurities can incorporate on either cation or anion sites, or by formation of \textit{DX} centers, in which a donor impurity exhibits a large lattice relaxation and becomes negatively charged, thus effectively acting as a deep acceptor.
The likelihood of \textit{DX}-center formation increases as the band gap increases, as is well known for AlGaAs~\cite{chadi1988theory} and AlGaN alloys~\cite{vandewalle1998,gordon2014hybrid}.

The problems encountered with Si indicate that investigating alternative shallow donors would be fruitful.
To date, incorporation of halide impurities (F and Cl) in {\ALGO} has not received much attention, in spite of the fact that they are widely recognized as effective dopants in transparent conducting oxides such as SnO$_2$~\cite{dixon2016}.
While they are expected to act as donors when incorporated on the oxygen site (i.e., F$_{\rm O}$ and Cl$_{\rm O}$)~\cite{varley2010oxygen,alfieri2021deep}, incorporation on other sites could potentially cause compensation~\cite{kang2017}.
Unintentional incorporation of these elements is indeed a concern since
during device processing, F and Cl are frequently introduced for etching purposes, for instance using hydrofluoric acid or by immersion in a fluorine-containing plasma~\cite{konishi20171}, or in Cl-based inductively coupled plasma reactive ion etching~\cite{shah2017inductively}.
Concerns about the impact of unintentionally incorporated F~\cite{yang2018effects} or Cl~\cite{alfieri2021deep} have been expressed.

In the present study, we use hybrid density functional theory to evaluate F and Cl impurities in Ga$_2$O$_3$ and Al$_2$O$_3$, considering both monoclinic and corundum phases. We assess not only their donor behavior on oxygen sites, but also possible compensation by interstitial configurations. 
We show that the halide interstitials act as compensating acceptors in all studied materials 
We also find, however, that the interstitials exhibit low migration barriers, offering promise to remove them in post-growth annealing.

For the substitutional sites, our results show that substitutional F and Cl on oxygen sites are effective shallow donors in both corundum and monoclinic Ga$_2$O$_3$. 
In wider-band-gap {\ALGO}, however, their behavior diverges.
In monoclinic {\ALGO}, F$_\mathrm{O(I)}$ [the O(I) site being the most stable site] forms compensating $DX$ centers already at an Al composition of 38\% Al.
At that concentration, the (+/0) charge-state transition level emerges into the band gap.
As we will show, donor ionization can still occur above that concentration, but with a reduction in the obtainable carrier concentration.
In contrast, Cl$_\mathrm{O(I)}$ does not exhibit $DX$ behavior, and remains a reliable shallow donor [for which the (+/0) level should be above the conduction-band minimum (CBM)] up to an Al composition of 50\% Al.
At that Al concentration, the (+/0) level drops below the CBM; donor ionization can still occur at higher Al concentrations, but with diminished doping efficiency.
For Cl$_\mathrm{O(II)}$ and Cl$_\mathrm{O(III)}$, the reduction in doping efficiency occurs at 36\% Al and 48\% Al, respectively.

In the corundum phase (sapphire), both F and Cl exhibit higher doping efficiency up to larger Al concentrations.
For F$_\mathrm{O}$, $DX$-center formation sets in only at 70\% Al. 
For Cl$_\mathrm{O}$, the result is particularly remarkable:
while Cl$_\mathrm{O}$ does undergo a $DX$ transition, this does not happen until the Al concentration reaches 84\%.
Even in pure corundum-phase $\alpha$-Al$_2$O$_3$, the ionization energy [determined as the position of the (+/--) transition level below the CBM] is only 0.48 eV, which could in principle yield observable $n$-type conductivity, particularly at elevated temperatures.  
This is a remarkable result for a material with a band gap of $\sim$9~eV, which would generally be considered an insulator.
Obtaining actual $n$-type conductivity will of course be challenging due to high formation energies and compensation by native defects as well as Cl interstitials.

The paper is organized as follows. In Sec.~\ref{method}, we discuss the methodology and calculation details. The main results for F and Cl are presented in Secs.~\ref{sec:monoclinic} (monoclinic) and \ref{sec:corundum} (corundum). 
Section~\ref{conc} concludes the paper. 

\section{Methodology} \label{method}
\begin{figure}
\includegraphics[width=0.47\textwidth]{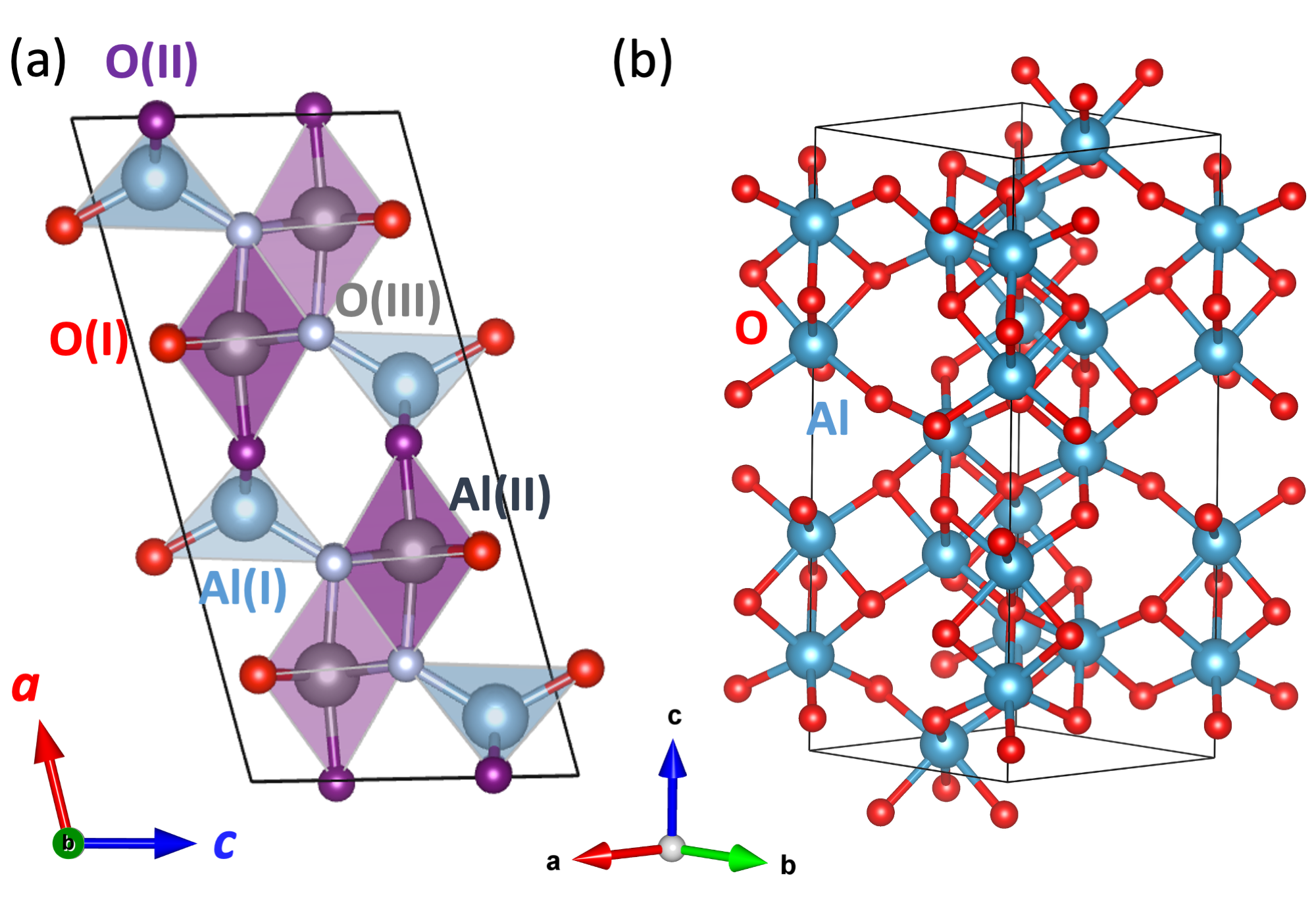}
\caption{\label{fig:str}
(a) Conventional cell of monoclinic {\AO} and (b) conventional cell of corundum {\AO}. In monoclinic {\AO}, inequivalent sites are labeled: tetrahedrally bonded Al(I) in aqua and octahedrally bonding Al(II) in purple, and threefold coordinated O(I) in red, threefold coordinated O(II) in purple and fourfold coordinated O(III) in gray.  In corundum {\AO}, there is only one inequivalent cation or anion site. Structural visualization was performed using VESTA~\cite{momma2011vesta}.
}
\end{figure}  

We performed DFT calculations using the projector augmented-wave method~\cite{Blochl1994} (with Ga $d$ states treated as part of the core), as implemented in the Vienna \textit{Ab-initio} Simulation Package (VASP).~\cite{Kresse1993,Kresse1996} Our study includes both the monoclinic phase (denoted as $\beta$ in the case of {\GO} and $\theta$ in the case of {\AO}) and the corundum phase ($\alpha$-{\GO} and {\AAO}); see Fig.~\ref{fig:str}.
Consistent with our previous calculations on impurities in the monoclinic phase~\cite{mu2022role} we used a plane-wave cutoff of 400~eV and the screened hybrid functional of Heyd–Scuseria–Ernzerhof (HSE)~\cite{heyd2003hybrid,*heyd2006erratum} with a mixing parameter $\alpha = 0.32$. 
For the corundum phase we relaxed the 10-atom rhombohedral primitive cell~\cite{mu2022phase} using a $\Gamma$-centered $6 \times 6 \times 6$ $k$-point mesh, and then constructed a hexagonal 30-atom conventional cell [see Fig.~\ref{fig:str}(b)] through the transformation matrix given in Sec.~I of  the Supplemental Material~\cite{Supp}.
We also constructed an orthorhombic 60-atom conventional cell based on the 30-atom conventional hexagonal cell with a transformation matrix given in the Supplemental Material~\cite{Supp}.
As seen in Table~\ref{tab:structure}, our calculated properties for both phases agree well with experiment.    

\begin{table*}
\caption{Structural parameters (lattice parameters, \AA; angle $\beta$, degrees), and band gaps (eV) for  {\GO} and {\AO}, and for monoclinic {\AAGO}. Experimental results are also listed for comparison.}
\begin{ruledtabular}
\begin{tabular}{ccccccccccc}
    & \multicolumn{2}{c}{\BGO} & \multicolumn{2}{c}{\AAGO}
    & \multicolumn{2}{c}{\TAO} & \multicolumn{2}{c}{\AGO} & \multicolumn{2}{c}{\AAO} \\
\hline
    & Calc & Expt & Calc & Expt & Calc & Expt & Calc & Expt & Calc & Expt \\
\hline
$a$      & 12.14$^a$ & 12.21$^b$
         & 11.86$^a$ & 12.00$^b$
         & 11.66$^a$ & 11.85$^c$
         & 4.92 & 4.98$^d$
         & 4.71 & 4.76$^e$   \\
$b$      & 3.02$^a$ & 3.04$^b$
         & 2.94$^a$ & 2.98$^b$
         & 2.88$^a$ & 2.90$^c$
         & --    & --
         & --    & --   \\
$c$      & 5.78$^a$ & 5.81$^b$
         & 5.69$^a$ & 5.73$^b$
         & 5.57$^a$ & 5.62$^c$
         & 13.24 & 13.43$^d$
         & 12.84 & 12.98$^e$  \\
$\beta$  & 103.77$^a$ & 103.87$^b$
         & 104.25$^a$ & 104.03$^b$
         & 104.04$^a$ & 103.83$^c$
         & --    & --
         & --    & --   \\
\hline
$E_\text{gap}$ & 4.83$^a$  & 4.76$^f$
               & 5.81$^a$  & --
               & 7.41$^a$ & 7.40$^g$
               & 5.60, 5.57$^h$ & 5.30$^i$
               & 9.09, 9.18$^j$ & 8.80$^k$
\end{tabular}
\end{ruledtabular}

{$^a$Ref.~\onlinecite{mu2022role};}
{$^b$Ref.~\onlinecite{kranert2015lattice};}
{$^c$Ref.~\onlinecite{zhou1991structures};}
{$^d$Ref.~\onlinecite{marezio1967bond};}
{$^e$Ref.~\onlinecite{lucht2003precise};}
{$^f$Ref.~\onlinecite{tippins1965optical,matsumoto1974absorption};}
{$^g$Ref.~\onlinecite{franchy1997growth};}
{$^h$Ref.~\onlinecite{lyons2019electronic};}
{$^i$Ref.~\onlinecite{shinohara2008heteroepitaxy};}
{$^j$Ref.~\onlinecite{choi2023};}
{$^k$Ref.~\onlinecite{innocenzi1990thermal,perevalov2007electronic,bortz1989optical};}
\label{tab:structure}
\end{table*}

For defect calculations we constructed 120-atom supercells as a $\mathrm{1\times2\times3}$ multiple of the 20-atom conventional unit cell for the monoclinic phase [Fig.~\ref{fig:str}(a)]~\cite{mu2022role}, with a $\mathrm{2\times2\times2}$ $k$-point grid; and a $\mathrm{2\times2\times1}$ multiple of the 30-atom conventional unit cell for the corundum phase [Fig.~\ref{fig:str}(b)]~\cite{mu2019influence}, with a single special $k$-point (1/4, 1/4, 1/4).
In addition, in order to ensure the accuracy of the obtained configurations, we performed checks in larger supercells.
For the monoclinic phase, we employed a 160-atom supercell, which is a $\mathrm{1\times2\times4}$ multiple of the 20-atom conventional cell, with a $\mathrm{2\times2\times2}$ $k$-point grid. 
For the corundum phase, we studied 270- and 360-atom supercells, which are $\mathrm{3\times3\times1}$ multiples of the 30-atom conventional hexagonal unit cell and $\mathrm{3\times2\times1}$ multiples of the 60-atom orthorhombic cell, respectively, using a single special $k$-point (1/4, 1/4, 1/4). 
We also checked $k$-point convergence: tests for Cl$_{\rm O}$ in corundum with a $\mathrm{2\times2\times2}$ $k$-point grid yielded charge-state transition levels within 0.01 eV of the results obtained using a single $k$-point.  
For the monoclinic phase, an ordered AlGaO$_3$ alloy was also investigated, with all Al atoms on octahedral sites and all Ga atoms on tetrahedral sites~\cite{peelaers2018structural,*peelaers2019erratum}.  
Spin polarization is included and full structural optimizations were performed.
Since $DX$-center configurations can be highly distorted, we used the ShakeNBreak code~\cite{mosquera2023identifying, mosquera2022shakenbreak} to generate multiple initial structures to ensure proper identification of the lowest-energy structure.    

Formation energies are calculated as follows, taking Cl$_\text{O}$ in {\AAO} as an example: 
\begin{equation}
\begin{split}
        E^f(\text{Cl}^q_\text{O}) & = E_\text{tot}(\text{Cl}^q_\text{O})-E_\text{tot}(\text{Al}_2\text{O}_3) -(\mu_\text{Cl}+\mu^0_\text{Cl}) \\
        & + (\mu_\text{O}+\mu^0_\text{O})+q(E_\text{F}+E_\text{VBM}) + \Delta^q ,
\end{split}
\end{equation}
where $E_\text{tot}(\text{Cl}^q_\text{O})$ is the total energy of one Cl$_\text{O}$ in charge state $q$ in the supercell, $E_\text{tot}(\text{Al}_2\text{O}_3)$ is the total energy of the bulk supercell, and $E_\text{F}$ is the Fermi energy, referenced to the valence-band maximum (VBM).
$\Delta^q$  is a finite-size correction term for charged defects~\cite{freysoldt2009fully, freysoldt2011electrostatic}.
The chemical potentials are referenced to the gas phases, i.e., $\mu^0_\text{Cl}=\frac{1}{2}E_\text{tot}(\text{Cl}_2)$ and $\mu^0_\text{O}=\frac{1}{2}E_\text{tot}(\text{O}_2)$.
For F, the chemical potential is bounded by formation of AlF$_3$; for Cl, by formation of AlCl$_3$ for Al-rich and Cl$_2$ for O-rich conditions.

Charge-state transition levels ($q/q'$) are defined as the Fermi-level position where the formation energies of charge states $q$ and $q'$ are equal.
The (+/0) and $(0/-)$ levels characterize the donor and acceptor nature of an impurity. In a $DX$ center, the $(0/-)$ level lies below the (+/0) level, characteristic of a ``negative-$U$'' center, where the effective correlation parameter $U$ is defined as the energy difference between the $(0/-)$ and the (+/0) levels~\cite{mu2022role}.
This implies that the neutral charge state is not thermodynamically stable, and hence the impurity is characterized by the ($+/-$) transition level, which we use as a descriptor: we consider the \textit{DX}-center is stable if the ($+/-$) level lies below the conduction-band minimum (CBM)~\cite{mu2022role}. 

Migration barriers are calculated using the climbing-image nudged elastic band (cNEB) method \cite{henkelman2000climbing}. To mitigate computational cost we perform one-shot HSE calculations for the migration barriers ($E_\text{b}$): we use the general gradient approximation of Perdew, Burke, and Ernzerhof (PBE)~\cite{Perdew1996} in the cNEB calculations, followed by static HSE total energy calculations based on the initial and barrier geometries.  The accuracy of this approach was tested in Ref.~\onlinecite{mu2022role}. 
Temperatures at which an impurity will be mobile are estimated based on harmonic transition state theory with an attempt frequency of 10$^{13}$ Hz and assuming a jump rate of one per second.

\section{Halides in the Monoclinic Phase} \label{sec:monoclinic}  

\begin{figure}
\includegraphics[width=0.52\textwidth]{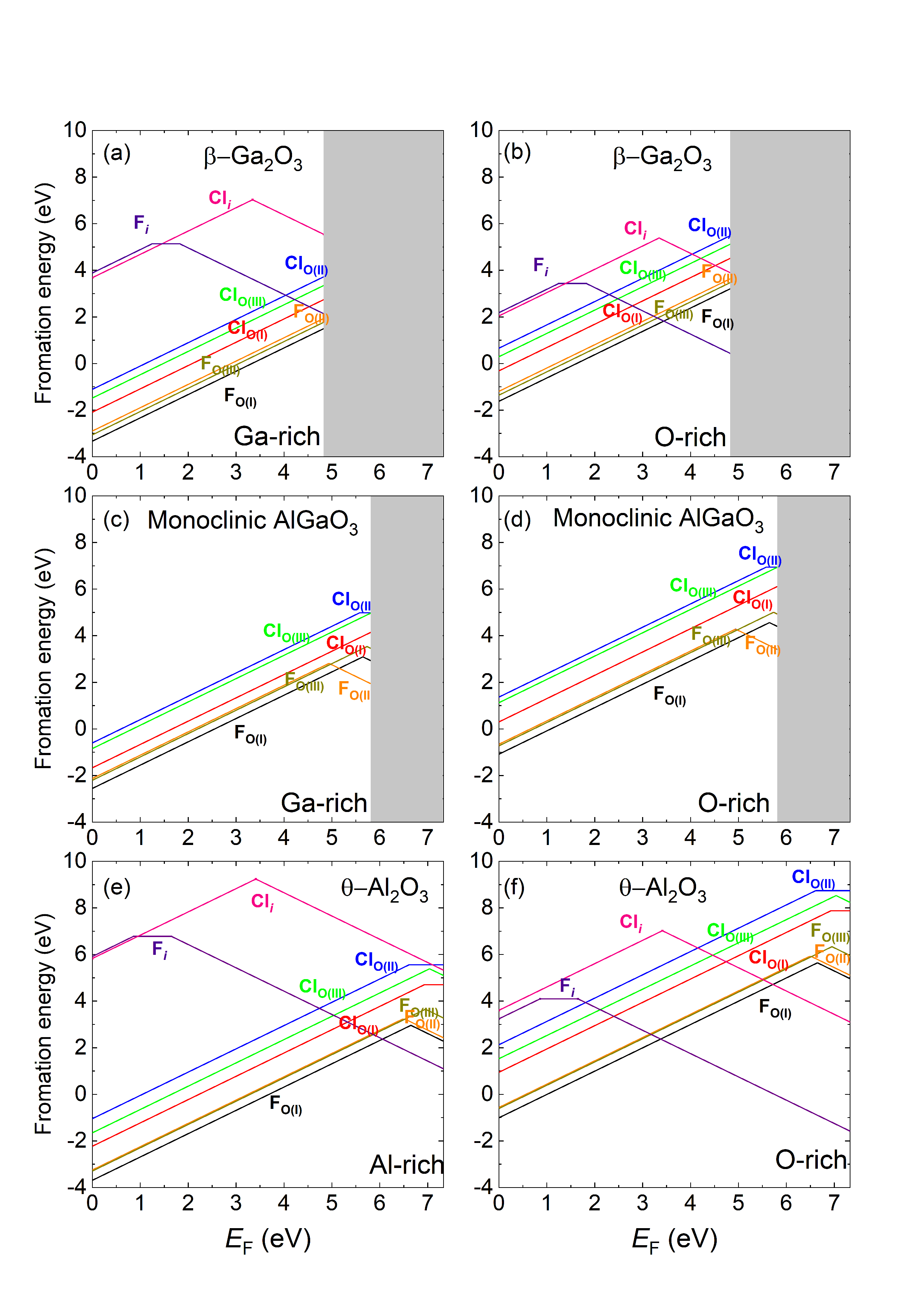}
\caption{\label{fig:formation}
Formation energy versus Fermi level for F and Cl impurities in (a)-(b) {\BGO}, (c)-(d) monoclinic AlGaO$_3$ and (e)-(f) {\TAO}.
Halide impurity incorporation on three possible O sites is considered: O(I), O(II), and O(III).
(a), (c) and (e) are for cation-rich, and (b), (d), (f) for O-rich conditions.
The grey area indicates the conduction band of {\BGO} and AlGaO$_3$.
}
\end{figure}  

\subsection{Fluorine} 
 
\subsubsection{F on oxygen sites, F$_\text{O}$} 

The formation  energies of F on three different O sites in monoclinic {\GO}, AlGaO$_3$ and {\AO} are shown in Fig.~\ref{fig:formation} for both cation-rich and O-rich conditions. 
In Fig.~\ref{fig:formation} (and all subsequent formation-energy figures) we keep the horizontal axis scale the same for {\BGO}, AlGaO$_3$, and {\TAO}, which facilitates comparisons of formation energies. We do not take the valence-band alignment between {\BGO} and {\TAO} into account since the valence-band offset is relatively small~\cite{peelaers2018structural,*peelaers2019erratum}. 

F$_{\rm O}$ in {\BGO} prefers the $\text{O(I)}$ site, consistent with the calculations in Ref.~\onlinecite{varley2010oxygen}, and only the positive charge state is stable for any Fermi level lying in the band gap; F$_\text{O}$ thus acts as a shallow donor in {\BGO}.  
The relative formation energies of F$_\text{O}$ follow the trend E$^f(\text{F}^+_\text{O(I)})$ $<$ E$^f(\text{F}^+_\text{O(III)})$ $<$ E$^f(\text{F}^+_\text{O(II)})$. 
The same trend holds for AlGaO$_3$ and {\TAO}. 

The formation energies for F$_\text{O}$ in monoclinic AlGaO$_3$ and {\TAO} are similar to those in {\BGO}.  
However, a distinct difference occurs when the Fermi level is high in the gap in the wider-band-gap materials: a $DX$ center forms in AlGaO$_3$ and {\TAO}.
When the Fermi level lies high in the gap, F$_\text{O}$ will be in the negative charge state; it therefore no longer acts as a shallow donor, but as a compensating acceptor. 
We note that a localized negative charge state can be stabilized in {\GO} as well, but results in a ($+/-$) charge-state transition level above the CBM of {\BGO} (Table~\ref{tab_sum}).

\begin{table}
\caption{Charge-state transition levels (eV) 
and effective correlation parameters $U$ (eV) for F$_\text{O}$, F$_i$, Cl$_\text{O}$ and Cl$_i$ in {\BGO}, monoclinic AlGaO$_3$, and {\TAO}. 
The neutral and negative charge states that are used to compute charge-state transition levels and $U$ all correspond to localized states.  
We also list $x^\text{onset}$ (\%), the Al concentration in {\ALGO} corresponding to the onset of \textit{DX} behavior [except for Cl$_\text{O(I)}$ and Cl$_\text{O(II)}$, where $x^\text{onset}$ corresponds to the concentration at which the (+/0) level moves below the CBM].  
For reference, the band gap of {\BGO} is 4.83~eV, for monoclinic AlGaO$_3$ 5.89~eV (based on the alloy bowing parameter), and for {\TAO} 7.41~eV.}
\begin{ruledtabular}
\begin{tabular}{cccccc}
{\BGO} &  ($+/0$)  & ($+/-$) 	& ($0/-$) 	& $U$ 	&	 \\
\hline
F$_\text{O(I)}$ &5.34  & 5.41 & 5.49 & 0.41 &   \\
F$_\text{O(II)}$ & 5.34 & 5.16 & 4.99& $-$0.35 &  \\
F$_\text{O(III)}$ & 5.31 & 5.70 & 6.09 & 0.79 &   \\
F$_i$ & 1.25 & 1.54 & 1.82 & 0.57 &   \\
Cl$_\text{O(I)}$ & 5.70 & 5.82  & 5.95 & 0.25  &  \\
Cl$_\text{O(II)}$ & 5.47 & 5.54  &5.61  & 0.14 &   \\
Cl$_\text{O(III)}$ & 5.94 & 5.73 & 5.52 & $-$0.42 &   \\
Cl$_i$ & 3.93 & 3.34 & 2.76 &  $-$1.16 &   \\
\hline
AlGaO$_3$ &  ($+/0$)  & ($+/-$) 	& ($0/-$) 	& $U$ 	&	 	 \\
\hline
F$_\text{O(I)}$ &5.66 & 5.65 & 5.63 & $-$0.03  &   \\
F$_\text{O(II)}$ & 5.10 & 4.94 & 4.79 & $-$0.30 &  \\
F$_\text{O(III)}$ & 5.83 & 5.74 & 5.64& $-$0.20  &   \\
Cl$_\text{O(I)}$ & 5.88 & 5.99  & 6.11 & 0.23  &  \\
Cl$_\text{O(II)}$ & 5.58 &6.15  &6.72  & 1.14 &   \\
Cl$_\text{O(III)}$ & 6.07 & 5.84 & 5.61 & $-$0.46 &   \\
\hline
{\TAO} &  ($+/0$)  & ($+/-$) 	& ($0/-$) 	& $U$ 	&	$x^\text{onset}$  	 \\
\hline
F$_\text{O(I)}$ &7.03 & 6.65 &6.27 & $-$0.76  & 38\%  \\
F$_\text{O(II)}$ & 6.47 & 6.50 & 6.53 & 0.05 & 15\%	  \\
F$_\text{O(III)}$ & 7.60 & 6.95 & 6.30 & $-$1.30  & 44\%  \\
F$_i$ & 0.86 & 1.26 & 1.65 & 0.79  &  -- \\
Cl$_\text{O(I)}$ & 6.93 & 7.18 & 7.44 & 0.50  & 50\% \\
Cl$_\text{O(II)}$ & 6.61 & 7.21 & 7.82 & 1.21  & 36\%	  \\
Cl$_\text{O(III)}$ & 7.38 & 7.04 & 6.70 & $-$0.68 & 48\%  \\
Cl$_i$ & 3.92 & 3.41 & 2.90 & $-$1.02  & --
\end{tabular}
\end{ruledtabular}
\label{tab_sum}
\end{table}

\begin{figure*}
\includegraphics[width=0.85\textwidth]{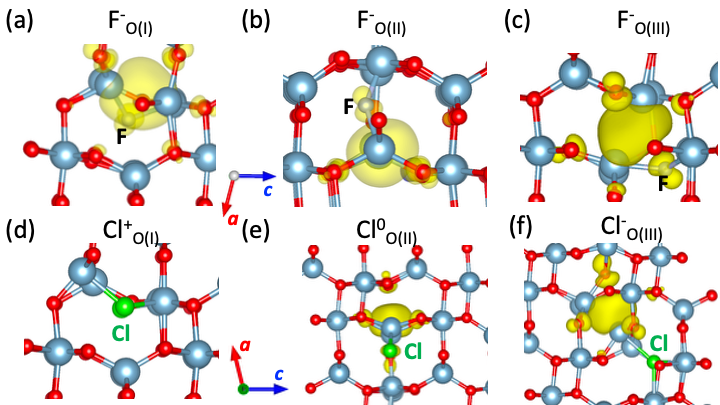}
\caption{\label{fig:chg}
Local structure of  F$_\text{O}$ and Cl$_\text{O}$ in {\TAO}: (a) F$^-_\text{O(I)}$, (b) F$^-_\text{O(II)}$, (c) F$^-_\text{O(III)}$, (d) Cl$^+_\text{O(I)}$, (e) Cl$^0_\text{O(II)}$, and (f) Cl$^-_\text{O(III)}$. Light blue spheres denote Al, red O, grey F, and green Cl.  Charge densities of the negatively-charged $DX$ configurations are shown in yellow, with isosurfaces of the charge density set at 0.005 $e$/\AA$^3$. 
}
\end{figure*} 

We now discuss the atomic configurations of the F$_\text{O}$ centers. 
In the positive charge state,
the F atom remains close to its ideal position without significant local distortions.
The bond length between F and the nearby Ga(I) site is approximately 1.93~\AA\ across all three F$_\text{O}$ cases, compared to the bulk O--Ga(I) bond lengths, which range from 1.83 to 1.86~\AA.
We attribute this to the closed-shell state of F$^+_\text{O}$ being analogous to that of the O anion, and their atomic sizes being similar as well.  

Atomic configurations for the negative charge state of F$_\text{O}$ (i.e., the $DX$ state) in {\TAO} are illustrated in Fig.~\ref{fig:chg}(a)--(c). 
F$_\text{O}$ still prefers O(I) sites, with a broken F--Ga(II) bond and a sizable atomic relaxation [Fig.~\ref{fig:chg}(a)].
In bulk {\TAO}, the O--Ga bond lengths are 1.86 and 1.86~\AA\ for two O(I)--Ga(II) bonds and 1.74~\AA\ for the O(I)-Ga(I) bond. 
In the bond-breaking configuration of F$^-_\text{O(I)}$, only two F-Ga bonds survive, with bond lengths of 1.85~\AA\ for F--Ga(I) and 1.93 \AA\ for F--Ga(II). 
This is a $DX$ configuration with a negative-$U$ value of $-$0.76 eV and a ($+/-$) level at 0.76~eV below the CBM (Table~\ref{tab_sum}). 
The charge density of $DX^-$ is localized around F$_\text{O(I)}$ [Fig.~\ref{fig:chg}(a)]. 

Similar comments apply to F$_\text{O(II)}$ and F$_\text{O(III)}$ in the negative charge state:
Strong local distortions are observed and charges are localized, as seen from Figs.~\ref{fig:chg}(b) and (c).  

The behavior of F$_\text{O}$ in ordered AlGaO$_3$ alloys is similar to the {\TAO} case. 
All F$_\text{O}$ are negative-$U$ centers with ($+/-$) transition levels lying below the CBM [see Fig.~\ref{fig:formation}(c)--(d) and Table~\ref{tab_sum}], and therefore they act as compensating acceptors when the Fermi level is close to the CBM. 

Given that F$_\text{O}$ acts as a shallow donor in {\BGO}, but becomes a deep level in AlGaO$_3$ and {\TAO}, it is interesting to figure out the Al concentration at which this transition occurs in {\ALGO} alloys. 
We therefore evaluate the position of the ($+/-$) transition level and compare it with the CBM in {\ALGO} alloys. 
As noted above, we were able to obtain a value for the ($+/-$) charge-state transition level in {\BGO}; along with the value for ordered AlGaO$_3$, we can perform a linear interpolation based on the values in Table~\ref{tab_sum}.
The values of the CBM are interpolated based on those in {\GO} and {\AO}, using a bowing parameter of 0.93 eV and assuming all of the band-gap bowing occurs in the CBM~\cite{peelaers2018structural,*peelaers2019erratum}.  
Using this bowing parameter, the bandgap of the 50\% alloy (AlGaO$_3$) is expected to be 5.89 eV, which is slightly larger than our DFT-calculated value of 5.81~eV in the ordered alloy.
The resulting values of $x^\text{onset}$, the Al concentration at which the ($+/-$) level moves below the CBM (i.e., the \textit{DX} center becomes stable) are listed in Table~\ref{tab_sum}.   
The key result is that F$_\mathrm{O}$ impurities start acting as $DX$-centers at modest Al concentrations.

\subsubsection{F on interstitial sites, F$_i$}  

\begin{figure}
\includegraphics[width=0.52\textwidth]{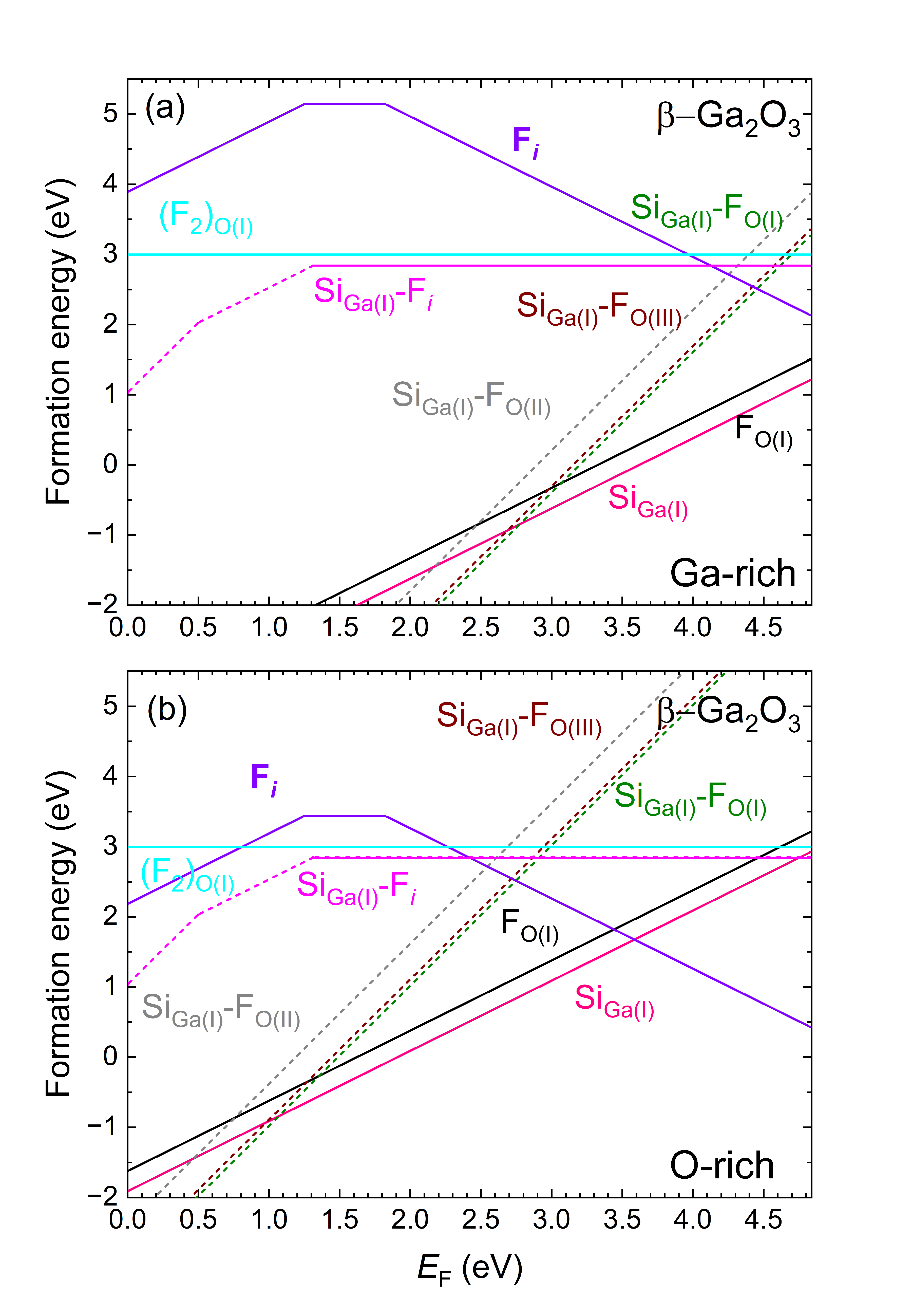}
\caption{\label{fig:complex}
Formation energy versus Fermi level for F$_\mathrm{O(I)}$, F$_i$, (F$_2$)$_\mathrm{O(I)}$, Si$_{\rm Ga(I)}$, and Si-F complexes in {\BGO} for (a) Ga-rich and (b) O-rich conditions. 
Dashed lines indicate a complex is not thermodynamically stable in a particular charge state. 
}
\end{figure}

Figure~\ref{fig:complex} shows the formation energy of the F interstitial (F$_i$) in {\BGO}, compared with F$_\mathrm{O(I)}$ (the lowest-energy configuration of substitutional F). 
F$_i$ is amphoteric, acting as a donor when the Fermi level is below 1.25~eV, and as an acceptor when the Fermi level is above 1.82 eV.  
The local geometry of F$^-_i$ is quite similar to H$^-_i$ in {\GO}~\cite{mu2022role}, with F bonded with two tetrahedral Ga sites. 
Interstitial fluorine can be stabilized further by bonding with F$_\mathrm{O}$, which leads to the (F$_2$)$_\mathrm{O(I)}$ complex in which two F atoms share a substitutional O site in a split-interstitial configuration.
(F$_2$)$_\mathrm{O(I)}$ is always charge neutral.
The attractive interaction between F$_i^-$ and F$_\mathrm{O(I)}^+$ leads to a binding energy of 0.65 eV, defined as the difference between the sum of the formation energies of the isolated defects and the formation energy of (F$_2$)$_\mathrm{O(I)}$.

Figure~\ref{fig:complex} shows that under both Ga-rich and O-rich conditions F$_i^-$ is lower in energy than (F$_2$)$_\mathrm{O(I)}$ when the Fermi level is closer to the CBM.
Under Ga-rich (O-poor) conditions, F$_\mathrm{O(I)}^+$ remains the lowest-energy configuration, but moving towards more O-rich conditions (which are more representative of growth), compensation by F$^-_i$ could clearly be significant, consistent with the experimental results of Ref.~\onlinecite{yang2018effects}.

Interstitials are usually quite mobile, and hence we investigate the migration barrier ($E_\text{b}$) of F$_i$ 
using the climbing-image nudged elastic band (cNEB) method~\cite{henkelman2000climbing} combined with a one-shot HSE approach.
We focus on the negative charge state, which is most relevant for $n$-type {\BGO}.
We find the migration barrier of F$^-_i$ in {\BGO} to be highly anisotropic.
The calculated values along the crystallographic directions are $E_\text{b}$([100]) = 1.73  eV;  $E_\text{b}$([010]) = 0.48 eV;  and $E_\text{b}$([001]) = 3.18 eV.  
The migration barrier along [010] is surprisingly low (0.48~eV) and indicates the interstitial can move well below room temperature.
It is thus unlikely to occur in isolated form, but will either diffuse out of the sample or bind to donors in the lattice.

Figures~\ref{fig:formation}(e)-(f) show the formation energy of F$_i$ in {\TAO} under Al-rich and O-rich conditions. 
The ($+$/$0$) and ($0$/$-$) charge-state transition levels occur at 0.86~eV and 1.65~eV above the VBM, respectively, corresponding to a positive $U$ value of 0.79~eV. 
F$_i$ exhibits similar qualitative behavior as in {\BGO}, i.e., to act as a compensating acceptor, with a positive $U$. 
While both F$_i$ and F$_\text{O}$ act as compensating acceptors for $n$-type {\TAO}, F$^-_i$ is more easily formed than F$^-_\text{O}$, even under Al-rich (O-poor) conditions [see Figs.~\ref{fig:formation}(e) and (f)]. 

Given the compensating nature of F$_i$ in both $n$-type {\BGO} and {\TAO} and its relatively low formation energy, we conclude that F$_i$, once unintentionally incorporated, poses a risk for compensation in monoclinic {\ALGO} alloys across the entire composition range. 
However, given the modest migration barriers reported above, these interstitials could potentially be removed in a post-growth anneal.

\subsubsection{Si--F complexes in~{\BGO}}  

Given that Si is widely used as a shallow donor in {\BGO} and {\ALGO} alloys, and that F may be unintentionally incorporated, we investigate the formation of Si--F complexes.  
We assume that Si resides on the tetrahedral Ga(I) site, which is energetically favored~\cite{mu2022role}.
F could reside on either a substitutional O site or an interstitial site.  

We first investigate Si$_\mathrm{Ga(I)}$--F$_\text{O}$ complexes.  Since both Si$_\mathrm{Ga(I)}$ and F$_\text{O}$ individually act as shallow donors in {\BGO}, we expect the complex to be a double donor. 
Indeed, all three Si$_\mathrm{Ga(I)}$--F$_\text{O}$ (with F incorporation on the three possible O sites) can occur only in the 2+ charge state. 
However, we find that they are not thermodynamically stable, based on the calculated binding energy of the complex:
\begin{equation}\label{eq:bind}
\begin{split}
E_\mathrm{bind}[(\mathrm{Si}_\mathrm{Ga(I)}-\mathrm{F_O})^{2+}] = E^f(\mathrm{Si}_\mathrm{Ga(I)}^+) +E^f(\mathrm{F}_\mathrm{O}^+) \\ 
- E^f[(\mathrm{Si}_\mathrm{Ga(I)}-\mathrm{F_O})^{2+}] \, .
\end {split}
\end{equation}
A positive value of the binding energy would indicate a stable, bound complex. 
However, we find negative binding energies for all three Si$_\mathrm{Ga(I)}$--F$_\text{O}$ complexes: --0.53~eV for O(I), --0.69~eV for O(II), and --0.35~eV for O(III). 
This implies that the separated constituents are more stable than the complex; we therefore indicate these complexes
with dashed lines in Fig.~\ref{fig:complex}. 

For the complexes with F$_i$, we find that for Fermi levels above 1.31~eV Si$_\mathrm{Ga(I)}$--F$_i$ complexes are stable in the neutral charge state, indicating that the Si donor has been passivated.  
We also find that Si$_\mathrm{Ga(I)}$--F$_i$ has (2$+/+$) and ($+/0$) charge-state transition levels at 0.50 and 1.31~eV above the VBM. However, these states are not thermodynamically stable due to negative binding energies:  $-$0.01 eV for (Si$_\mathrm{Ga(I)}$--F$_i$)$^+$ and $-$0.21 eV for (Si$_\mathrm{Ga(I)}$--F$_i$)$^{2+}$. We use dashed lines to illustrate this instability in Fig.~\ref{fig:complex}. 

The binding energy of (Si$_\mathrm{Ga(I)}$--F$_i$)$^0$ is found to be 0.50 eV. 
We can estimate an activation energy for dissociation by adding the migration barrier of F$_i$ to the binding energy.
Using the lowest calculated migration barrier for F$^-_i$, we obtain 0.50 + 0.48 = 0.98~eV, indicating that the complex would easily dissociate at or slightly above room temperature. 
We conclude that complex formation between F and Si is unlikely to play any role in actual materials.

\subsection{Cl impurity}  

\subsubsection{Cl on oxygen sites, Cl$_\text{O}$} \label{clo}

Formation  energies for Cl on different O sites are shown in Fig.~\ref{fig:formation} for both cation-rich and O-rich conditions. 
Cl$_{\rm O}$ in {\BGO} has been investigated previously~\cite{varley2010oxygen,alfieri2021deep} and our results are consistent with those calculations:  
only the positive charge state is stable for any Fermi level lying in the band gap, and thus Cl$_\text{O}$ (on any of the three O sites) acts as a shallow donor in {\BGO}.  
The relative formation energies of Cl$_\text{O}$ follow the same trend as F$_\text{O}$, that is,  E$^f(\text{Cl}^+_\text{O(I)})$ $<$ E$^f(\text{Cl}^+_\text{O(III)})$ $<$ E$^f(\text{Cl}^+_\text{O(II)})$. 
This trend holds for AlGaO$_3$ and {\TAO}.  
The charge-state transition levels of all three Cl$_\text{O}$ sites are well above the CBM, as listed in Table~\ref{tab_sum}.

The results for Cl$_\text{O}$ in {\TAO} are generally similar to those in {\BGO}: the formation energies have similar magnitudes, and Cl$_\text{O}$ prefers the $\text{O(I)}$ site.
The formation energy of Cl$_\text{O(I)}^+$ in {\TAO} is lower than that of Cl$_\text{O(III)}$ (Cl$_\text{O(II)}$) by 0.58 (1.18) eV. 
Overall, the formation energies of Cl$_\text{O}$ are somewhat higher than those for F$_\text{O}$.

The charge-state transition levels are summarized in Table~\ref{tab_sum}.
All the results  for formation energies and charge-state transition levels for Cl$_{\text{O}}$ in {\TAO} were verified using a larger 160-atom supercell.  
In {\TAO} ($E_g=$7.41~eV), Cl$_{\text{O(I)}}$ exhibits a ($+$/0) charge-state transition level at 6.93~eV above the VBM (see Table~\ref{tab_sum}), corresponding to 0.48~eV below the CBM.  
The negative charge state of Cl$_{\text{O(I)}}$ in {\TAO} can also be stabilized; it involves significant local distortions and charge localization around the Cl atom and nearby Al atoms. The resulting (0/$-$) charge-state transition level lies 0.03~eV above the CBM (Table~\ref{tab_sum}). As a result, Cl$_{\text{O(I)}}$ is a  positive-$U$ center with $U$=0.50~eV. 

The position of the (+/0) transition level in {\TAO} precludes Cl$_{\text{O(I)}}$ from acting as an effective shallow donor.  
However, with an ionization energy of 0.48~eV, we expect that observable $n$-type conductivity could occur.
Based on a model of carrier statistics (see Supplemental Material~\cite{Supp}) that uses the effective density of states of the conduction band $N_c(\mathrm{T}) = 6.3\times10^{18}(\mathrm{T/300K})^{3/2} \ \mathrm{cm^{-3}}$ and a Cl concentration of $10^{19} \ \mathrm{cm^{-3}}$, we find a carrier concentration of 
$\mathrm{5 \times 10^{14} \ cm^{-3}}$ at 300~K and $\mathrm{9 \times 10^{16} \ cm^{-3}}$ at 600~K.

To estimate the onset Al concentration at which Cl$_{\text{O(I)}}$ becomes neutral in monoclinic {\ALGO} alloys, we interpolate the ($+$/0) transition level between AlGaO$_3$ and {\TAO} and compare it with the position of the CBM in {\ALGO}. 
We find that the Cl$_{\text{O(I)}}$ (+/0) levels moves below the CBM at approximately 50\% Al composition.

Looking at incorporation of Cl on other O sites in {\TAO}, we find that
Cl$_{\text{O(II)}}$ exhibits a ($+/0$) charge-state transition level within the band gap, at 0.80~eV below the CBM, with a positive $U$ value of 1.21~eV. 
Interpolating the ($+$/0) transition level between {\BGO} and AlGaO$_3$, we find that Cl$_{\text{O(II)}}$ remains an effective donor up to approximately 36\% Al composition.

Cl$_{\text{O(III)}}$ exhibits a ($+/-$) charge-state transition level within the band gap, i.e., it behaves as a DX center, with a negative $U$ value of $-$0.68 eV. 
However, the position of this ($+/-$) level at 0.37 eV below the CBM suggests that it can still be relatively easily ionized. 
Based on the carrier-statistics model outlined in the Supplemental Material~\cite{Supp}, we find that this value of the (+/$-$) level enables a carrier concentration that will saturate at $\mathrm{4 \times 10^{12} \ cm^{-3}}$ at 300~K and $\mathrm{1 \times 10^{16} \ cm^{-3}}$ at 600~K.
Interpolating the ($+/-$) charge-state transition level of Cl$_{\text{O(III)}}$ between AlGaO$_3$ and {\BGO}, we find that the onset of self-compensation occurs at 48\% Al composition.

Selected atomic configurations for Cl$_\text{O}$ in {\TAO} are illustrated in Fig.~\ref{fig:chg}(d)--(f). 
In the positive charge state [Fig.~\ref{fig:chg}(d)], strong outward displacements of the nearest neighbors are observed, resulting in an increase of the Cl--Al bond length from the bulk value of 1.86 {\AA}  to 2.22--2.28 {\AA} for Cl$^+_\text{O(I)}$ and from the bulk value of 1.98 {\AA}  to 2.24--2.39 \AA\ for Cl$^+_\text{O(III)}$. 
We attribute this to the large atomic size of Cl relative to oxygen.

The atomic configuration for the neutral charge state of Cl$_\text{O(II)}$ [Fig.~\ref{fig:chg}(e)] is noteworthy. 
Three nearest-neighbor Al atoms are pushed away from the Cl center, leading to an increase in the  Al--Cl bond lengths from the bulk values of 1.73, 1.73, 1.85 \AA\ to 2.21, 2.18, 2.21 {\AA}.  
The electron charge is localized between two Al(I), one of which is a nearest neighbor of Cl [Fig.~\ref{fig:chg}(e)], and the distance between these Al(I) atoms is reduced from 2.88 {\AA} to 2.36~{\AA}.

Figure~\ref{fig:chg}(f) shows the local structure of Cl$^{-}_{\mathrm{O(III)}}$.   
The nearest-neighbor Al(I) atom is pushed outward, increasing the Al(I)--Cl bond length from 1.79~\AA{} to 2.46~\AA, and 
the two  electrons end up being confined between two Al(I) sites. 
A nearby O(III) atom is displaced from its ideal lattice position toward an interstitial site.

\subsubsection{Cl on interstitial sites, Cl$_\textrm{i}$} \label{cli} 

Calculations for Cl$_i$ in {\BGO} were reported in Ref.~\onlinecite{alfieri2021deep}.
A pronounced negative-$U$ behavior was observed with the (+/--)\ charge-state transition level located 3.3~eV above the VBM. 
Our calculations yield a very similar value, 3.34~eV (Table~\ref{tab_sum}).
Cl$_i$ thus acts as a compensating acceptor in $n$-type {\BGO}.  
However, Ref.~\onlinecite{alfieri2021deep} found that formation energy of Cl$_i$ to be higher than that of Cl$_\text{O}$, even under O-rich conditions.
In our calculations [Fig.~\ref{fig:formation}(a)-(b)], this is true only under Ga-rich conditions, and under O-rich conditions the formation energy of Cl$_i$ becomes lower than that of Cl$_\text{O(I)}$ by 0.56 eV at the CBM. 
This difference may be due to the different treatment of Ga-$d$ states: Ref.~\onlinecite{alfieri2021deep} included Ga $d$ states as valence states, while 
here we treat Ga $d$ states treated as part of the core. 

Figures~\ref{fig:formation}(e)-(f) show the formation energy of Cl$_i$ in {\TAO} under Al-rich and O-rich conditions.   
Similar to Cl$_i$ in {\BGO}, Cl$_i$ in {\TAO} is a negative-$U$ center (Table~\ref{tab_sum}) with a (+/$-$) charge-state transition level at 3.41~eV above the VBM. 
Cl$_i$ will thus act as a compensating acceptor. 
Under Al-rich conditions, Cl$_i$ has a higher formation energy than Cl$_\text{O}$ for $n$-type materials, while its formation energy becomes notably lower than Cl$_\text{O}$, posing greater risk for compensation. 
Similar to F$_i$, Cl$_i$, once unintentionally incorporated, thus acts as a compensating acceptor in {\ALGO} alloys across the entire alloying composition.

We also investigate the direction-dependent migration barriers ($E_\text{b}$) of Cl$_i$ in {\BGO}
using the climbing-image nudged elastic band (cNEB) method~\cite{henkelman2000climbing} combined with a one-shot HSE approach.
We focus on the negative charge state (Cl$^-_i$), which is most relevant for $n$-type {\BGO}.
The calculated values along the crystallographic directions are $E_\text{b}$([100]) = 3.26  eV;  $E_\text{b}$([010]) = 0.26 eV;  and $E_\text{b}$([001]) = 4.50 eV. 
The migration barrier of Cl$^-_i$ in {\BGO} along [010] is  surprisingly low 0.26~eV, even lower than the migration barrier of F$^-_i$. We also calculated the migration barrier of Cl$^-_i$ in {\TAO} along [010] and found it to be 0.31~eV. 
This indicates a very small migration barrier across all compositions of {\ALGO}. 
The interstitial is thus unlikely to occur in the isolated form, but will either diffuse out of the sample or bind to donors in the lattice of {\ALGO}.

\section{Halides in the Corundum Phase} \label{sec:corundum}  

\subsection{Fluorine} 

\subsubsection{F on oxygen sites, F$_\text{O}$} 

\begin{figure}
\includegraphics[width=0.52\textwidth]{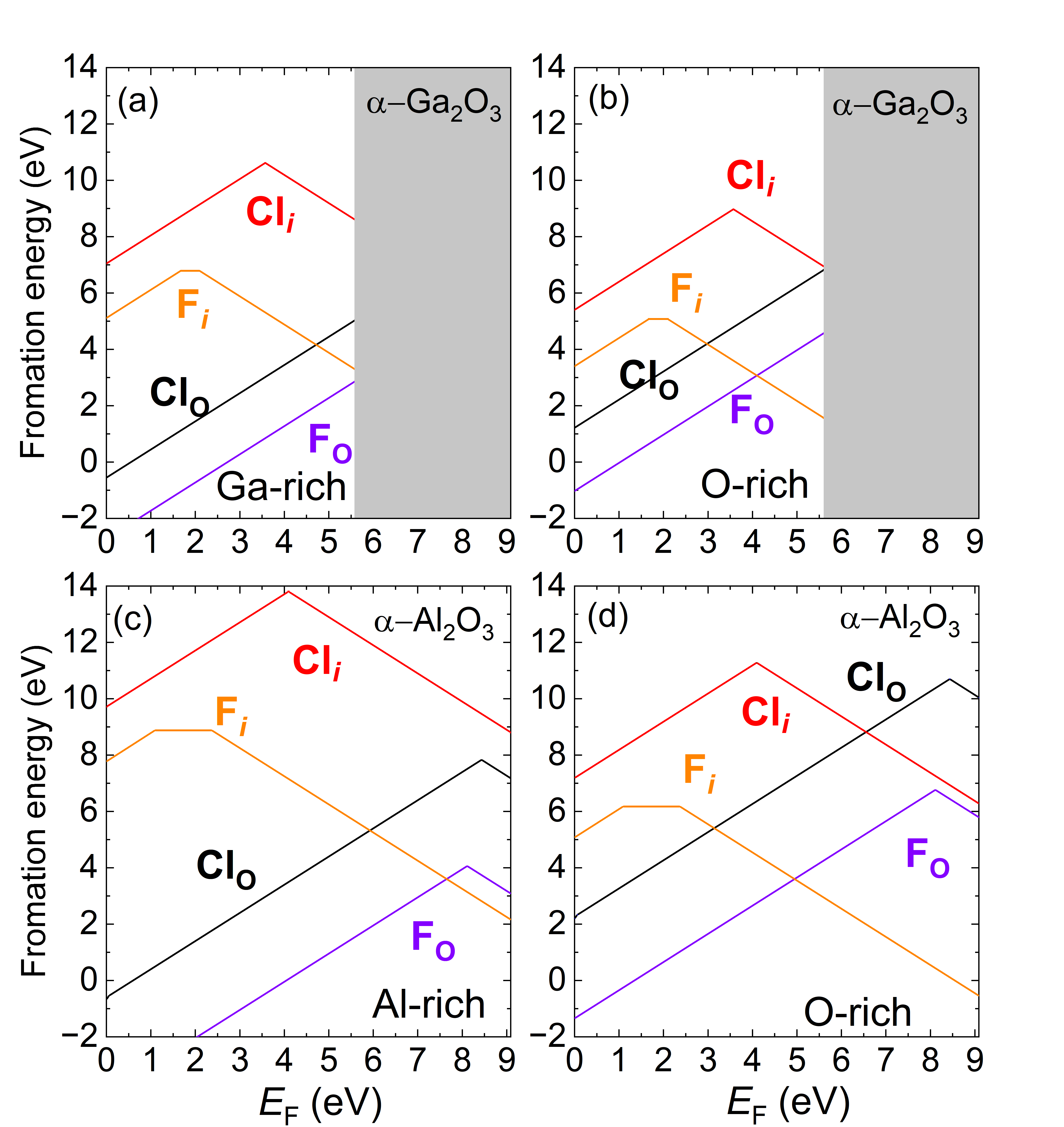}
\caption{\label{fig:formsapphire}
Formation energy versus Fermi level for F and Cl impurities in (a)-(b) {\AGO} and (c)-(d) {\AAO}.
(a) and (c)  are for cation-rich, and (b) and (d) for O-rich conditions.
}
\end{figure}

Figure~\ref{fig:formsapphire}(a)--(b) shows the formation energy of fluorine in the corundum phase of {\GO} ({\AGO}).  
Even though the band gap of {\AGO} (5.60~eV) is higher than that of {\BGO} (4.83~eV), F$_{\rm O}$ in {\AGO} remains an effective shallow donor, similar to its behavior in {\BGO}. 
This is consistent with recent experiments on F-doped {\AGO}~\cite{choi2025fluorine}.
F$^+_{\rm O}$ is more easily formed under Ga-rich (O-poor) conditions, and its formation energy at the VBM is comparable to that in {\BGO}. 
A localized F$^-_{\rm O}$ state can be stabilized, but it has a high formation energy and results in a (+/--)\ transition level located 1.0~eV above the CBM.

Fluorine in the corundum phase of {\AO} ($\alpha$-Al$_2$O$_3$) was recently investigated in detail with DFT-HSE calculations~\cite{choi2023}. 
For completeness, our own results for formation energies of F in {\AAO} are included in Fig.~\ref{fig:formsapphire}(c)--(d).
The results are similar to our findings for F$_{\rm O}$  in {\TAO}:
for Fermi-level positions high in the band gap, a negatively charged $DX$ configuration is lower in energy.
The local geometries of F$_\mathrm{O}$ in {\AGO} are similar to those in {\AAO}; the latter were described in Ref.~\onlinecite{choi2023}.

The ($+/-$) charge-state transition levels in corundum {\ALGO} alloys are interpolated based on the values in {\AGO} and {\AAO} and listed in Table~\ref{tab_corundum}). 
The values of the CBM are interpolated based on those in {\AGO} and {\AAO} using a bowing parameter of 1.78 eV~\cite{peelaers2018structural} and assuming all of the band-gap bowing occurs in the CBM. 
For F$_{\rm O}$, the ($+/-$) level moves below the CBM at an Al concentration of 70\%; the resulting value of $x^\mathrm{onset}$ is also listed in Table~\ref{tab_corundum}.

\begin{table}
\caption{ 
Charge-state transition levels (eV) and effective correlation parameters $U$ (eV) for F$_\text{O}$, F$_i$, Cl$_\text{O}$ and Cl$_i$ in {\AGO} and {\AAO}. 
The neutral and negative charge states that are used to compute charge-state transition levels and $U$ all correspond to localized states;
We also list $x^\text{onset}$ (\%), the Al concentration in {\ALGO} corresponding to the onset of \textit{DX} behavior.
For reference, the band gap of {\AGO} is 5.60~eV and for {\AAO} 9.09~eV.
}
\begin{ruledtabular}
\begin{tabular}{cccccc}
{\AGO} &  ($+/0$)  & ($+/-$) 	& ($0/-$) 	& $U$ & 	 \\
\hline
F$_\text{O}$ &6.66  & 6.64 & 6.63 & $-$0.03 &  \\
F$_i$ &1.67  & 1.88 & 2.09 & 0.41 &    \\
Cl$_\text{O}$ &7.22  & 6.62 & 6.02 & $-$1.20 &   \\
Cl$_i$ &3.63  & 3.57 & 3.51 & -0.13  &  \\
\hline
 {\AAO} &  ($+/0$)  & ($+/-$) 	& ($0/-$) 	& $U$  & $x^\text{onset}$		 	 \\
 \hline
 F$_\text{O}$ &9.41  & 8.11 & 6.82 & $-$2.60 & 70\%  \\
F$_i$ &1.36  & 1.87 & 2.37 & 1.01   & --- \\
Cl$_\text{O}$ &9.12  & 8.61 & 8.10 & $-$1.02 &  84\%  \\
Cl$_i$ &4.09  & 4.10 & 4.10 & 0.01 &  --- \\
\end{tabular}
\end{ruledtabular}
\label{tab_corundum}
\end{table}

\subsubsection{F on interstitial sites, F$_i$}  

Fluorine interstitials in {\AGO} display  ($+$/0) and (0/$-$) charge-state transition levels located at 1.67 and 2.09~eV above the VBM. 
F$_i$ in {\AGO} thus acts as a compensating center when $E_\text{F}$ lies above 2.09~eV, and the energy of F$_i^-$ in $n$-type material is lower than that of F$_{\rm O}$, except under extreme Ga-rich conditions.
Since its formation energy is low, compensation by F$_i^-$ is thus a genuine risk.

For fluorine interstitials in {\AAO}, we again find agreement in charge-state transition levels, formation energies, and local geometries with Ref.~\onlinecite{choi2023}, and similar behavior as for F$_i$ in {\TAO}.
The atomic configuration of F$^-_i$, as described in Ref.~\onlinecite{choi2023}, is very similar to our calculated local geometry of F$^-_i$ in {\AGO} and {\AAO}.
In $n$-type {\AAO}, F$_i^-$ is lower in energy than F$_{\rm O}^-$ [Fig.~\ref{fig:formsapphire}(c)--(d)], and thus poses an even higher risk of compensation.

Our NEB calculations show that the migration barrier of F$^-_i$ in {\AAO} is 2.65~eV along the [001] direction and 3.47~eV along the [100] (or the symmetry-equivalent [010]) direction.    
While the barrier of 2.65 eV is higher than in the monoclinic phase, F$^-_i$ would still be mobile at temperatures above about 700$^\circ$C, indicating that annealing could be used to remove F$^-_i$ acceptors and thereby reduce compensation.

\subsection{Chlorine} \label{Cl:sapphire}   

\subsubsection{Cl on oxygen sites, Cl$_\text{O}$} \label{clo4}

The results for Cl$_\text{O}$ in {\AGO} [Fig.~\ref{fig:formsapphire}(a)--(b)] are very similar to those in {\BGO} [Fig.~\ref{fig:formation}(a)--(b)]: the formation energies have similar magnitudes, and Cl$_\text{O}$ also acts as a shallow donor.
Note that there is only one inequivalent O site in the corundum structure.
For Cl$^-_\text{O}$, we were able to find a locally stable localized configuration, 
but the (+/--) transition level was found to be well above the CBM (by 1.02~eV; see Table~\ref{tab_corundum}).

In Sec.~\ref{clo} we found that Cl$_\text{O(I)}$ and Cl$_\text{O(III)}$ may lead to observable carrier concentrations in monoclinic {\AO}. 
This raises the question whether Cl$_\text{O}$ could still act as a donor in the corundum phase, which has an even larger band gap (9.09~eV in our calculations) than monoclinic {\TAO} (7.41~eV).

\begin{figure*}
\includegraphics[width=0.95\textwidth]{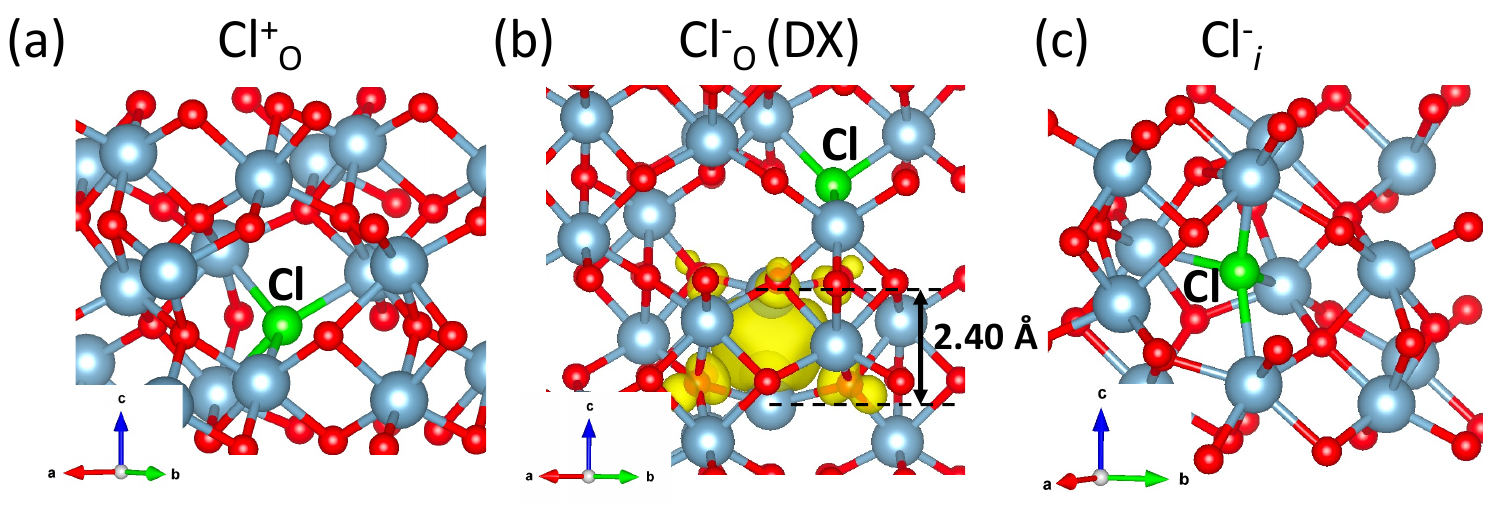}
\caption{\label{fig:cl-sapphire}
Local structure of Cl$_\text{O}$ and Cl$_i$ in {\AAO} (sapphire): (a) Cl$^+_\text{O}$, (b) Cl$^-_\text{O}$ (DX), (c) Cl$^-_i$. Light blue spheres denote Al, red O, and green Cl.   
The charge density of the negatively charged $DX$ configuration is shown in yellow, with the isosurface value set at 0.006 $e$/\AA$^3$. 
} 
\end{figure*} 

Figure~\ref{fig:formsapphire}(c)--(d) shows the formation energies of Cl$_\text{O}$ in {\AAO} for Al-rich and O-rich conditions. 
Cl$_\text{O}$ is found to be a negative-$U$ center, with $U = -1.02$~eV, and 
a ($+$/$-$) charge-state transition level at 0.48~eV below the CBM (Table~\ref{tab_corundum}).

The local geometry is shown in Fig.~\ref{fig:cl-sapphire}(a).  
For the positive charge state, we find that Cl pushes the nearest-neighbor Al atoms outward, leading to elongated Cl--Al bond lengths ranging from 2.11 to 2.23~\AA, compared to the bulk O--Al bond lengths of 1.84 and 1.95~\AA.
These values are comparable to the Cl--Al bond lengths for Cl$_\text{O}$ in monoclinic {\AO} (see Sec.~\ref{clo}). 

The geometry of the negative charge state (i.e., the $DX$ configuration) is shown in Fig.~\ref{fig:cl-sapphire}(b).  
One of the four Al-Cl bonds is broken, pushing that Al atom away from Cl, primarily along the $c$ axis. 
Just beneath this Al atom (along the $c$ axis), another Al atom moves towards it, forming a Al-Al dimer with a bond length of 2.40 {\AA}. 
This is significantly smaller than the original bond distance between the two Al atoms (3.83~{\AA}). 
The electrons that form the negative charge state are confined within the Al-Al dimer, as illustrated by the isosurface in Fig.~\ref{fig:cl-sapphire}(b). 
A similar local geometry is observed in the ground state of Cl$^{0}_\text{O}$, with the electron also localized within the dimer. 

The large size of these relaxations prompted us to perform checks in larger supercells, including a 270-atom and a 360-atom supercell. 
Consistent results were obtained for all three supercell sizes. 

By interpolating the Cl$_\mathrm{O}$ ($+/-$) charge-state transition levels between the end compounds ({\AGO} and {\AAO}), we found that the Al concentration at which the (+/$-$) level moves below the CBM 
in corundum {\ALGO} is $x^\mathrm{onset}$ = 84\%.

\subsubsection{Cl on interstitial sites, Cl$_\textrm{i}$} \label{cli} 

Cl$_i$ in {\AGO} exhibits negative-$U$ behavior, with the ($+/-$)\ charge-state transition level located at 3.57~eV above the VBM. Above this level, Cl$_i$ acts as a compensating acceptor. Under both Ga-rich and O-rich conditions, the formation energy of Cl$_i$ is higher than that of Cl$_\text{O}$, and the overall behavior is qualitatively consistent with that of Cl$_i$ in {\BGO}.

Interstitial Cl$_i$ in {\AAO} is amphoteric; the ($+/0$) and ($0/-$) charge-state transition levels lie very close together, at 4.09 and 4.10~eV above the VBM [Fig.~\ref{fig:formsapphire}(c)-(d)].
Cl$_i$ will thus act as an acceptor when the Fermi level is above midgap. 
The atomic structure of Cl$^-_i$ places the Cl atom near the inversion center of $\alpha$-{\AO}, occupying a tetrahedral interstitial site coordinated by four nearby Al atoms [Fig.~\ref{fig:cl-sapphire}(c)]  
  
Our NEB calculations show that the migration barrier of Cl$^-_i$ in {\AAO} is 2.75~eV along the [001] direction and 2.73 ~eV along [100] (or the symmetry-equivalent [010]) direction. 
Cl$^-_i$ would thus still be mobile at temperatures above about 700$^\circ$C, indicating that annealing could be used to remove Cl$^-_i$ acceptors and thereby reduce compensation.

\subsubsection{$n$-type doping of {\AO}?} \label{ntype} 

We put a question mark in the heading of this section because achieving actual $n$-type doping of a material with as large a band gap as {\AO} (7.41~eV for {\TAO}, 9.09~eV for {\AAO}) faces many obstacles---some of which will be addressed here.

The first prerequisite is finding an impurity that acts as a shallow donor.   

Among the donor impurities that have been investigated in prior~\cite{wickramaratne2022,choi2013,choi2023} and present work, Cl is most promising.
We found that Cl$_\text{O}$ in {\TAO} has a ($+/0$) charge-state transition level located 0.48~eV below the CBM, and Cl$_\text{O}$ in {\AAO} has a ($+/-$) charge-state transition level at 0.48~eV below the CBM. 
Although this level is not particularly ``shallow'' in the conventional semiconductor sense, {\AAO} would typically be regarded as an insulator, and finding a dopant that has a transition level so close to the CBM is remarkable.

The position of the transition level determines the electron concentration $n$ in semiconductor devices, and it is informative to
estimate the electron concentration that could be obtained in {\AO} using Cl as a donor, in the absence of compensation.
Using the methodology outlined in Sec.~II of the Supplemental Material~\cite{Supp}, and assuming an impurity concentration $N_\mathrm{Cl} = 10^{19}\ \mathrm{cm^{-3}}$, we obtain a carrier concentration at 300~K of 
$\mathrm{5 \times 10^{14} \ cm^{-3}}$ in {\TAO} and $\mathrm{6 \times 10^{10} \ cm^{-3}}$ in {\AAO}.
However, since {\AO} is a ceramic, high-temperature applications might be envisioned.
At 600~K, the carrier concentrations would increase to 
$\mathrm{9 \times 10^{16} \ cm^{-3}}$ in {\TAO} and $\mathrm{2 \times 10^{15} \ cm^{-3}}$ in {\AAO}.
Note that in the case of the $DX$ center (i.e., in {\AAO}) the carrier concentration saturates once $N_\mathrm{Cl}$ exceeds the electron concentration corresponding to the Fermi-level position being pinned near the (+/$-$) level, with a value of $2 \times10^{15}\ \mathrm{cm^{-3}}$ at 600~K [see the Supplemental Material~\cite{Supp}].
In the absence of other compensation, increasing $N_\mathrm{Cl}$ above $10^{16}\ \mathrm{cm^{-3}}$ would therefore not result in additional carriers.

Finding a donor with a sufficiently shallow level is a necessary but not sufficient conditions to achieve true $n$-type conductivity.
The dopant also needs to be incorporated in sufficiently high concentrations.
As shown in Figs.~\ref{fig:formation} and \ref{fig:formsapphire}, the formation energy of Cl$^{+}_\text{O}$ is high,
suggesting that the expected equilibrium concentration of Cl$_\text{O}$ is low. 
However, dopant impurities are commonly incorporated under non-equilibrium conditions (such as implantation followed by annealing).

We focused on the lowest-energy configuration of Cl in {\TAO}, namely Cl on the O(I) site. 
In Sec.~\ref{clo}, we found that Cl on the O(III) site exhibits a $(+/-)$ charge-state transition level 0.37~eV below the CBM, which is closer to the CBM than the $(+/0)$ level of Cl$_{\mathrm{O(I)}}$. 
This suggests that Cl$_{\mathrm{O(III)}}$ can still be relatively easily ionized, yielding an electron concentration of $1 \times10^{16}\ \mathrm{cm^{-3}}$ at 600~K, provided
the concentration of Cl on the $\mathrm{O(III)}$ site exceeeds that value.
However, the formation energy of Cl$^{+}_{\mathrm{O(III)}}$ in {\TAO} is 0.58~eV higher than that of Cl$_{\mathrm{O(I)}}$, making this configuration unlikely to form under equilibrium conditions. 
Nevertheless, non-equilibrium incorporation could lead to a finite concentration of Cl$_{\mathrm{O(III)}}$.

A more important problem is compensation.  While unintentional incorporation of acceptor impurities could hopefully be controlled, formation of Cl interstitials and native point defects is much harder to avoid.  
Regarding Cl$^-_i$, annealing could potentially remove them from the sample.
Aluminum vacancies ($V_{\rm Al}$) will be the main problem, since they occur in a 3$-$ charge state in $n$-type material and their formation energy will be quite low, even under Al-rich conditions~\cite{choi2023, choi2013}.
If $V_{\rm Al}$ formation can be suppressed under non-equilibrium conditions, the ability for Cl to drive the Fermi very high in {\AAO} could be demonstrated.

Even if full $n$-type conditions cannot be achieved, driving the Fermi level close to the CBM in a material that is typically considered an insulator would be very exciting.
In addition, identifying an experimental technique that could confirm the presence and character of a shallow donor (even when compensated by acceptors) would be highly interesting; perhaps some variation of photothermal spectroscopy~\cite{Kogan1967} could be employed.

\section{Conclusion}\label{conc}
In summary, we have conducted a comprehensive first-principles study of F and Cl impurities in Ga$_2$O$_3$ and Al$_2$O$_3$, considering both monoclinic and corundum phases. 
Fluorine easily incorporates on the oxygen site, and acts as a shallow donor in {\GO}, but is prone to self-compensation through $DX$ center formation in {\ALGO} alloys. 
We find that
F$_\mathrm{O}$ turns into a compensating acceptor at 38\% Al in monoclinic
and 70\% Al in corundum {\ALGO} alloys. 
Fluorine interstitials act as compensating acceptors in all studied materials, but
exhibit low enough migration barriers to be removed by annealing.
We also studied the interaction between F and Si, the most widely used donor in {\BGO}.
A neutral Si$_\mathrm{Ga}$-F$_{i}$ complex may form, but the calculated activation energy for dissociation (0.98~eV) indicates that the complex would easily dissociate during annealing.

Chlorine shares many of the characteristics of fluorine; for instance, Cl$_i$ interstitials act as deep acceptors in all studied $n$-type materials---but they also have modest migration barriers.
The main difference with fluorine lies in the fact that substitutional chlorine exhibits more favorable donor behavior than fluorine. 
Cl$_\mathrm{O}$ acts as a shallow donor in Ga$_2$O$_3$.
Its doping efficiency starts to decrease in {\ALGO} alloys at 50\% Al for the lowest-energy Cl$_\mathrm{O(I)}$ in monoclinic {\ALGO} (due to an increase in the ionization energy), and at 84\% Al in corundum {\ALGO} (due to $DX$ center formation).  

In pure {\AO}, Cl$_\mathrm{O}$ is not strictly speaking a shallow donor, but its transition level lies only 0.48~eV below the CBM (in both the monoclinic and corundum phase), indicating it could be ionized and lead to measurable carrier concentrations at modest temperatures.
These results identify Cl as a highly promising shallow donor in both monoclinic and corundum {\ALGO} alloys.
 
At the same time, achieving true $n$-type conductivity in pure {\AAO} (sapphire) would be highly challenging. The calculated formation energy of Cl$_\mathrm{O}$ is high, and compensation by native acceptors, particularly aluminum vacancies, will be difficult to avoid. 
Non-equilibrium incorporation methods could be tried to demonstrate the donor behavior of Cl$_\mathrm{O}$. 
Even if full $n$-type conductivity cannot be achieved, experimental observation of a relatively shallow donor in a material that is traditionally considered an insulator would   
be highly exciting.

\begin{acknowledgments} 

The authors acknowledge Xuecong Wang for generating initial structures with local distortions.  
The work was supported by the Air Force Office of Scientific Research (FA9550-18-1-0479 and FA9550-22-1-0165).
Use was made of computational facilities purchased with funds from the National Science Foundation (CNS-1725797) and administered by the Center for Scientific Computing (CSC). The CSC is supported by the California NanoSystems Institute and the Materials Research Science and Engineering Center (MRSEC; NSF DMR 2308708) at UC Santa Barbara.
We also used the Expanse supercomputer at the San Diego Supercomputer Center through allocations PHY230093 and DMR070069 from the Advanced Cyberinfrastructure Coordination Ecosystem: Services \& Support (ACCESS) program, which is supported by National Science Foundation Grants No.2138259, No. 2138286, No. 2138307, No. 2137603, and No. 2138296. 
This research also used resources of the National Energy Research Scientific Computing Center, a DOE Office of Science User Facility supported by the Office of Science of the U.S. Department of Energy under Contract No. DE-AC02-05CH11231 using NERSC award BES-ERCAP0028497.

\end{acknowledgments}

\appendix

\nocite{Medvedeva2007Electronic}
\nocite{Neufeld2016Play}
\bibliography{Cl_F_070824_cleaned}
\end{document}